\documentclass[preprint,floatfix,aps,prd,amsmath,amssymb,nofootinbib,preprintnumbers,superscriptaddress]{revtex4}

\usepackage{graphicx}
\usepackage{array}

\begin{document}

\begin{flushright}
ICRR-Report-589-2011-6\\
\end{flushright}

\title{Higher Order Corrections to the Primordial Gravitational Wave
  Spectrum and its Impact on Parameter Estimates for Inflation}

\author{Sachiko Kuroyanagi} \email[]{skuro@icrr.u-tokyo.ac.jp}
\affiliation{Institute for Cosmic Ray Research, University of Tokyo,
  Chiba 277-8582, Japan}

\author{Tomo Takahashi} 
\affiliation{Department of Physics, Saga University, Saga 840-8502, Japan}

\begin{abstract}
  We study the impact of the use of the power series expression for
  the primordial tensor spectrum on parameter estimation from future
  direct detection gravitational wave experiments.  The spectrum
  approximated by the power series expansion may give large deviation
  from the true (fiducial) value when it is normalized at CMB scale
  because of the large separation between CMB and direct detection
  scales.  We derive the coefficients of the higher order terms of the
  expansion up to the sixth order within the framework of the
  slow-roll approximation and investigate how well the inclusion of
  higher order terms improves the analytic prediction of the spectrum
  amplitude by comparing with numerical results.  Using the power
  series expression, we consider future constraints on inflationary
  parameters expected from direct detection experiments of the
  inflationary gravitational wave background and show that the
  truncation of the higher order terms can lead to incorrect
  evaluation of the parameters.  We present two example models; a
  quadratic chaotic inflation model and mixed inflaton and curvaton
  model with a quartic inflaton potential.
\end{abstract}

\maketitle

%%%%%%%%%%%%%%%%%%%%%%%%%%%%%%%%
%%%%%%%%%%%%%%%%%%%%%%%%%%%%%%%%
\section{Introduction}
%%%%%%%%%%%%%%%%%%%%%%%%%%%%%%%%
%%%%%%%%%%%%%%%%%%%%%%%%%%%%%%%%
Inflation \cite{inflation} is a successful paradigm not only for
solving the horizon and flatness problems, but also for explaining the
origin of density perturbations in the Universe.  Inflation predicts
adiabatic and almost scale-invariant primordial fluctuations, which
are in excellent agreement with current observations such as cosmic
microwave background (CMB) and so on.  However, no direct evidence of
inflation has yet been found.  During inflation, the gravitational
waves could also be produced \cite{staro}, whose detection can give
a direct evidence of inflation and would be a key test of inflation.

Early detection of the inflationary gravitational wave background may
be achieved through its unique signature in the polarization of the
CMB \cite{CMBp1,CMBp2}.  The ongoing satellite mission, Planck
\cite{:2006uk}, can detect such indirect signal of gravitational waves
if the tensor-to-scalar ratio is $r\gtrsim 0.05$.  The next-generation
experiment, such as CMBpol \cite{Baumann:2008aq} and Cosmic Origins
Explorer (COrE) \cite{Collaboration:2011ck}, are designed to reach $r
\sim 10^{-3}$.  Moreover, the direct detection may be possible with
space-based laser interferometers such as the DECi-hertz
Interferometer Gravitational wave Observatory (DECIGO)
\cite{Seto:2001qf,Kawamura:2006up} and Big-Bang Observer (BBO)
\cite{bbo}, which would provide independent information about
inflation.

While CMB polarization experiments observe large-scale gravitational
waves ($k\sim \mathcal{O}(0.001)~{\rm Mpc}^{-1}$), space-based laser
interferometers measure gravitational waves at millihertz frequencies
($k\sim \mathcal{O}(10^{13})~{\rm Mpc}^{-1}$).  This millihertz
frequency band is the most prospective region for direct detection of
the inflationary gravitational wave background.  The detection becomes
easier at lower frequencies, since interferometer with longer arms can
obtain larger displacement signals by gravitational waves.  On the
other hand, frequencies below a millihertz would be contaminated by
the gravitational wave background generated from white dwarf binaries
\cite{Farmer:2003pa}.

We should note that there are also many other mechanisms which may
generate a gravitational wave background around the millihertz
frequency, such as preheating
\cite{Khlebnikov:1997di,preheatGW1,preheatGW2,preheatGW3}, bubble
collisions during a first-order phase transition
\cite{phasetransition}, self-ordering scalar fields following a global
phase transition \cite{Krauss:1991qu,Fenu:2009qf}, second order
effects from enhanced scalar perturbations
\cite{Saito:2008jc,Assadullahi:2009nf,Jedamzik:2010hq,Suyama:2011pu},
topological defects
\cite{Vilenkin:1981bx,Caldwell:1991jj,Damour:2000wa,Kawasaki:2010yi,Olmez:2010bi,Gleiser:1998na,Hiramatsu:2010yz},
supernova explosions of population III first stars
\cite{Buonanno:2004tp, Sandick:2006sm, Suwa:2007du}, gamma-ray bursts
\cite{Suwa:2009si}, and so on.  However, their amplitude and frequency
strongly depend on their unknown physics, so the millihertz band is
still a window to search for the inflationary gravitational wave
background.  In this paper, we focus on the gravitational wave
background from inflation and do not consider other sources which may
contaminate the millihertz band.

The large difference between CMB and direct detection scales means
that these two types of observations enable us to look at different
periods of inflation, which would greatly help to investigate the
inflaton potential
\cite{Ungarelli:2005qb,Smith:2005mm,Smith:2006xf,Smith:2008pf,
  Kuroyanagi:2009br,Easson:2010zy,Easson:2010uw}.  However, we should
carefully choose the method to connect the two separate scales.  A
common method is to use a power-law extrapolation from CMB scales to
direct detection scales for describing the primordial tensor spectrum.
Yet recent works
\cite{Chongchitnan:2006pe,Friedman:2006zt,Kuroyanagi:2008ye} have
pointed out that the Taylor expansion around the CMB scale is no
longer valid at the direct detection frequency and it causes an
incorrect estimation of the amplitude of the inflationary
gravitational wave background.

One way to avoid the wrong estimation of the spectrum is to resort to
a full numerical calculation to obtain the gravitational wave
spectrum.  However, the power-law extrapolation is much simpler and
easier than the numerical method and, in principle, its precision can
be improved by including higher order terms in the Taylor expansion as
much as possible.  In this paper, we derive the slow-roll expression
for the primordial tensor power spectrum up to the sixth order in the
Taylor expansion and examine how much the inclusion of the higher
order terms improves the estimation of the amplitude at direct
detection scales by comparing with the full numerical computation
\cite{Kuroyanagi:2008ye}.  Furthermore, we discuss the impact of the
truncation of the higher order terms in the power series expansion of
the tensor spectrum by presenting constraints on inflationary
parameters expected from future direct detection experiments, which is
an example that such a poor estimation of the spectrum amplitude
causes a problem.

This paper is organized as follows: In Sec.~\ref{slow-roll}, we give a
formula of the power series expression for the primordial tensor
spectrum including up to the sixth order in the Taylor expansion.
Next, in Sec. \ref{overestimation}, we discuss whether the expression
given in Sec.~\ref{slow-roll} well describes the tensor power spectrum
by comparing those with numerically obtained spectra.  We consider two
example models for the comparison, the chaotic inflation model with
quadratic and quartic potentials.  Although the quartic chaotic
inflation is already excluded by observations such as CMB, by adding
the contribution from another source of fluctuations such as the
curvaton, the quartic inflation model can be allowed due to the
existence of the curvaton fluctuations, which is sometimes called
mixed inflaton and curvaton model
\cite{Langlois:2004nn,Moroi:2005kz,Moroi:2005np,Ichikawa:2008iq}.
Note that this kind of mixed scenario can give sizable
tensor-to-scalar ratio as well as large non-Gaussianity, which might
be interesting from the viewpoint of near future observations.  In
Sec. \ref{parameterestimate}, we give expected constraints on the
inflationary parameters for the above mentioned two models. In
passing, we discuss to what extent the truncation of the tensor
spectrum expression at some (lower) order leads to incorrect
evaluation of the inflationary parameters.  Finally, we conclude in
Sec. \ref{conclusion}.

%%%%%%%%%%%%%%%%%%%%%%%%%%%%%%%%
%%%%%%%%%%%%%%%%%%%%%%%%%%%%%%%%
\section{Slow-roll formalism and power series expansion}
\label{slow-roll}
%%%%%%%%%%%%%%%%%%%%%%%%%%%%%%%%
%%%%%%%%%%%%%%%%%%%%%%%%%%%%%%%%
In the standard picture of the early universe, a scalar field $\phi$,
the inflaton, drives superluminal cosmic expansion, the inflation.
The equation of motion for $\phi$ is given by
\begin{equation}
\ddot{\phi}+3H\dot{\phi}+V^{\prime}(\phi)=0,
\end{equation}
where the dot and prime denote the derivative with respect to $t$ and
$\phi$, respectively.  The dynamics of inflation is often
characterized by the slow-roll parameters.  In this paper, we work
with the slow-roll parameters which are defined in terms of the
inflaton potential $V$ and its derivatives as \cite{Liddle:1994dx}
\begin{equation}
\begin{array}{rcl}
&&\epsilon_V\equiv \displaystyle\frac{M_{\rm Pl}^2}{2}\left(\displaystyle\frac{V^{\prime}}{V}\right)^2, \vspace{3mm}\\ 
&&\eta_V\equiv M_{\rm Pl}^2 \displaystyle\frac{V^{\prime\prime}}{V},\vspace{3mm}\\
&&\xi^2_V\equiv M_{\rm Pl}^4 \displaystyle\frac{V^{\prime}V^{\prime\prime\prime}}{V^2},\vspace{3mm}\\
&&\sigma^3_V\equiv M_{\rm Pl}^6 \displaystyle\frac{V^{\prime 2}V^{(4)}}{V^3},\vspace{3mm}\\
&&\tau^4_V\equiv M_{\rm Pl}^8\displaystyle\frac{V^{\prime 3}V^{(5)}}{V^4},\vspace{3mm}\\
&&\zeta^5_V\equiv M_{\rm Pl}^{10}\displaystyle\frac{V^{\prime 4}V^{(6)}}{V^5},
\end{array}
\end{equation}
where the subscript $(n)$ denotes the $n$-th derivative with respect
to $\phi$.  Inflation lasts as long as $\epsilon_V,|\eta_V|\ll 1$,
which are called the slow-roll conditions, and it ends when this
condition is violated, $\max\{ \epsilon_V(\phi_{\rm end}),
\eta_V(\phi_{\rm end})\}=1$.  In the slow-roll limit, the evolution of
the Hubble parameter $H(t)$ is given by $H^2\simeq V/(3M_{\rm Pl}^2)$,
where $M_{\rm Pl}=1/\sqrt{8\pi G}$ is the reduced Planck mass.  The
duration of inflation is characterized by the e-folding number,
$N\equiv \ln(a_{\rm end}/a)$, which can be rewritten in terms of the
potential,
\begin{equation}
N \simeq \frac{1}{M_{\rm Pl}^2}\int^{\phi}_{\phi_{\rm end}}
\frac{V}{V^{\prime}}d\phi.
\label{eqN}
\end{equation}

Within the slow-roll approximation, the primordial power spectra of
scalar and tensor perturbations are given by
\cite{Stewart:1993bc,Lidsey:1995np}
\begin{eqnarray}
&&{\cal P}_{S}\simeq \left.[1-(2C+1)\epsilon_H+C\eta_H]^2\frac{1}{16\pi^2M_{\rm Pl}^4}\frac{H^4}{H^{\prime 2}}\right|_{k=aH},\\
&&{\cal P}_{T}\simeq [1-(C+1)\epsilon_H]^2\frac{2}{\pi^2 M_{\rm Pl}^2}H^2|_{k=aH},
\label{eqPT2nd}
\end{eqnarray}
where $C=-2+\ln 2+\gamma\simeq -0.73$ with $\gamma$ being the Euler
constant and $\epsilon_H$ and $\eta_H$ are the Hubble slow-roll
parameters, $\epsilon_H\equiv 2M_{\rm Pl}^2(H^{\prime}/H)^2$ and
$\eta_H\equiv 2M_{\rm Pl}^2H^{\prime\prime}/H$.  Hereafter, we only
consider the leading order for the slow-roll parameters.  Then the
power spectra are given in terms of the inflaton potential as
\begin{eqnarray}
&&{\cal P}_{S}\simeq \left.\frac{1}{12\pi^2 M_{\rm Pl}^6}\frac{V^3}{V^{\prime 2}}\right|_{k=aH},
\label{eqPS}\\
&&{\cal P}_{T}\simeq \frac{2}{3\pi^2 M_{\rm Pl}^4}V|_{k=aH}.
\label{eqPT}
\end{eqnarray}
They are evaluated at the moment when each Fourier mode $k$ crosses
the Hubble horizon, as indicated by the subscript ``$k=aH$.''  It is
often assumed that the values of $V$ and its derivatives evolve so
slowly during inflation that the spectra can be parametrized by using
the Taylor expansion in terms of the logarithm of the wave number,
\begin{eqnarray}
{\cal P}_T(k)= {\cal P}_{T\star}
\exp\left[n_{T\star}\ln\frac{k}{k_\star}
+\frac{1}{2!}\alpha_{T\star}\ln^2\frac{k}{k_\star}
+\frac{1}{3!}\beta_{T\star}\ln^3\frac{k}{k_\star}\right.\nonumber\\
\left.
+\frac{1}{4!}\gamma_{T\star}\ln^4\frac{k}{k_\star}
+\frac{1}{5!}\delta_{T\star}\ln^5\frac{k}{k_\star}
+\frac{1}{6!}\theta_{T\star}\ln^6\frac{k}{k_\star}+\cdots\right],
\label{primordial}
\end{eqnarray}
where the coefficients are the parameters characterizing a deviation
from the scale-invariant spectrum,
\begin{equation}
\begin{array}{rcl}
n_T(k)\equiv &&\displaystyle\frac{d\ln {\cal P}_T(k)}{d\ln k},\vspace{3mm} \\ 
\alpha_T(k)\equiv &&\displaystyle\frac{dn_T(k)}{d\ln k},\vspace{3mm} \\
\beta_T(k)\equiv &&\displaystyle\frac{d\alpha_T(k)}{d\ln k},\vspace{3mm}  \\ 
\gamma_T(k)\equiv &&\displaystyle\frac{d\beta_T(k)}{d\ln k},\vspace{3mm} \\
\delta_T(k)\equiv &&\displaystyle\frac{d\gamma_T(k)}{d\ln k},\vspace{3mm} \\
\theta_T(k)\equiv &&\displaystyle\frac{d\delta_T(k)}{d\ln k}.
\end{array}
\end{equation}
The expression for the scalar power spectrum is the same except that
the coefficient of the first term is $(n_{S\star}-1)$.  The subscript
$\star$ denotes quantities evaluated at the pivot scale, which is
commonly taken to be the scale of the CMB, $k_\star=0.002{\rm
  Mpc}^{-1}$.  The coefficients can be given in terms of the slow-roll
parameters as
\begin{equation}
\begin{array}{rcl}
n_T(k)\simeq &&-2\epsilon_V\\
\alpha_T(k)\simeq &&-4\epsilon_V[2\epsilon_V-\eta_V],\\
\beta_T(k)\simeq &&-4\epsilon_V[16\epsilon_V^2+2\eta_V^2-14\epsilon_V\eta_V+\xi^2_V],\\
\gamma_T(k)\simeq &&-4\epsilon_V[192\epsilon_V^3-236\epsilon_V^2\eta_V+72\epsilon_V\eta_V^2-4\eta_V^3+22\epsilon_V\xi^2_V-7\eta_V\xi^2_V-\sigma^3_V],\\
\delta_T(k)\simeq &&-4\epsilon_V[3042\epsilon_V^4-4810\epsilon_V^3\eta_V+2280\epsilon_V^2\eta_V^2-328\epsilon_V\eta_V^3+8\eta_V^4\\
&&+500\epsilon_V^2\xi^2_V-324\epsilon_V\eta_V\xi^2_V+33\eta_V^2\xi^2_V+7\xi_V^4
-32\epsilon_V\sigma^3_V+11\eta_V\sigma^3_V+\tau_V^4],\\
\theta_T(k)\simeq &&-4\epsilon_V[61440\epsilon_V^5-117840\epsilon_V^4\eta_V+75200\epsilon_V^3\eta_V^2-18272\epsilon_V^2\eta_V^3+1408\epsilon_V\eta_V^4-16\eta_V^5\\
&&+12840\epsilon_V^3\xi^2_V-12596\epsilon_V^2\eta_V\xi^2_V+3000\epsilon_V\eta_V^2\xi^2_V-131\eta_V^3\xi^2_V+408\epsilon_V\xi_V^4-94\eta_V\xi_V^4\\
&&-948\epsilon_V^2\sigma^3_V+648\epsilon_V\eta_V\sigma^3_V-77\eta_V^2\sigma^3_V-25\xi_V^2\sigma^3_V+44\epsilon_V\tau_V^4-16\eta_V\tau_V^4-\zeta^5_V].
\label{coeff}
\end{array}
\end{equation}
The amplitude of the tensor perturbation at the CMB scale is often
parametrized by the tensor-to-scalar ratio:
\begin{equation}
r\equiv \frac{{\cal P}_{T\star}}{{\cal P}_{S\star}}=16\epsilon_{V\star}.
\end{equation}

As we will show in the next section, the inclusion of the higher order
terms in the Taylor expansion up to the 6th order seems to be in very
good agreement with a full numerical calculation, which indicates that
the above expressions would be precise enough to give correct tensor
power spectra for many inflation models and can be used for parameter
estimation from observational data.

%%%%%%%%%%%%%%%%%%%%%%%%%%%%%%%%
%%%%%%%%%%%%%%%%%%%%%%%%%%%%%%%%
\section{Overestimation of the tensor power spectrum}
\label{overestimation}
\begin{figure}
 \begin{center}
  \includegraphics[width=0.6\textwidth]{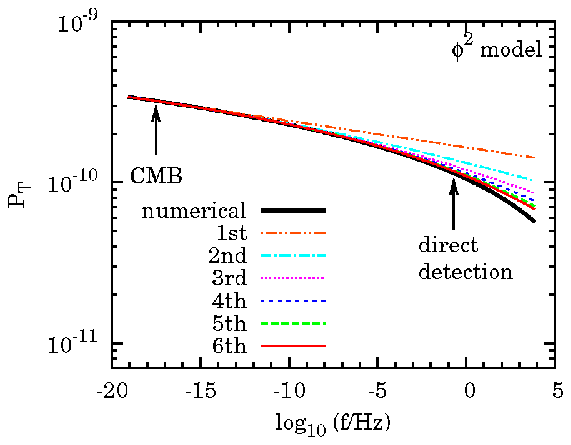}
  \includegraphics[width=0.6\textwidth]{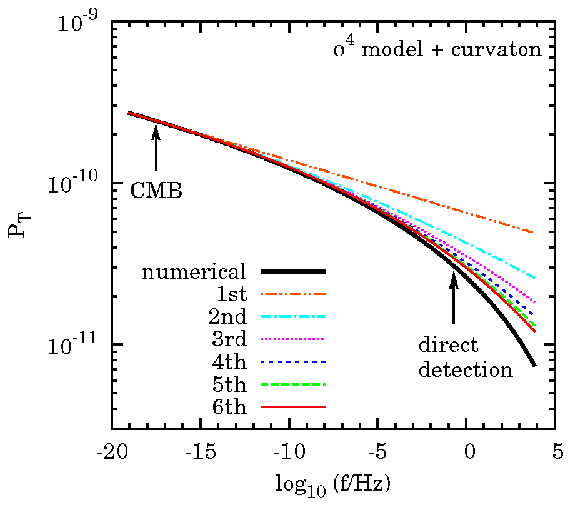}
  \caption{\label{spectrum} Comparison between the exact
    (numerically obtained) spectrum and the spectra approximated by
    truncating the Taylor expansion after the first, second, third,
    fourth, fifth or sixth order terms in Eq. (\ref{primordial}).  The
    primordial tensor spectra ${\cal P}_T$ are plotted against
    frequency, $f=k/2\pi$.  The upper panel shows the case with the
    quadratic chaotic inflation model, $V=m^2\phi^2/2$.  The bottom
    panel shows the case for a mixed inflaton and curvaton model with
    the quartic inflaton potential, $V=\lambda\phi^4/4$. The fraction
    of the curvaton contribution is fixed requiring that the
    tensor-to-scalar ratio is $r=0.1$. }
 \end{center}
\end{figure}

%%%%%%%%%%%%%%%%%%%%%%%%%%%%%%%%
%%%%%%%%%%%%%%%%%%%%%%%%%%%%%%%%
In most works, it is common to simply adopt the power-law
extrapolation from CMB scales to direct detection scales for
describing the gravitational wave background spectrum.  However, as we
will show below, such a power-law extrapolation may not be valid and
lead to an incorrect estimation of the spectrum amplitude at direct
detection scales.  In Figure~\ref{spectrum}, we show the gravitational
wave spectra calculated using the Taylor expansion truncating at some
order and the one obtained from full numerical computations.  Here we
consider quadratic ($\phi^2$) and quartic ($\phi^4$) chaotic inflation
models.  For the $\phi^4$ model, we consider a mixed inflaton and
curvaton model where fluctuations from the curvaton
\cite{Enqvist:2001zp,Lyth:2001nq,Moroi:2001ct} also contribute to
cosmic density perturbations.  This is because the quartic chaotic
inflation model predicts too large tensor-to-scalar ratio which is
already excluded by current observational data.  In addition, the
curvaton model can generate large non-Gaussianity, thus such a mixed
model would be interesting to investigate since it can produce both
sizable gravitational wave amplitude and large non-Gaussianity
detectable in the near future observations \footnote{
  Mixed inflaton-curvaton models have been studied in
  \cite{Langlois:2004nn,Moroi:2005kz,Moroi:2005np,Ichikawa:2008iq} and
  we refer the readers to these papers for details.
}.

As seen from the figure, the power series expression overestimates the
amplitude of the spectrum because of the large separation between the
two scales.  The spectra are plotted using Eq.~(\ref{primordial}) by
truncating the Taylor expansion at each order, respectively.  The
exact spectrum, which is obtained from a numerical calculation
\cite{Kuroyanagi:2008ye}, is also plotted for comparison.  The
truncation of the higher order terms in Eq.~(\ref{primordial}) is the
cause of the overestimation because the contribution of the higher
order terms is non-negligible as they are boosted by the $n$-th power
of $\ln(k_{0.2{\rm Hz}}/k_\star)\simeq 38.7$, even though the
coefficients of the $n$-th terms are suppressed as $\epsilon^n$.  The
overestimation of the spectrum amplitude can be avoided if the
slow-roll parameters are much smaller than $[\ln(k_{0.2{\rm
    Hz}}/k_\star)]^{-1}\simeq 2.58 \times 10^{-2}$, but this is not
the case, in particular, for chaotic inflation models.

%%%%%%%%%%%%%%%%%%%%
%%%%%%%%%%%%%%%%%%%%

\begin{figure}
 \begin{center}
  \includegraphics[width=0.7\textwidth]{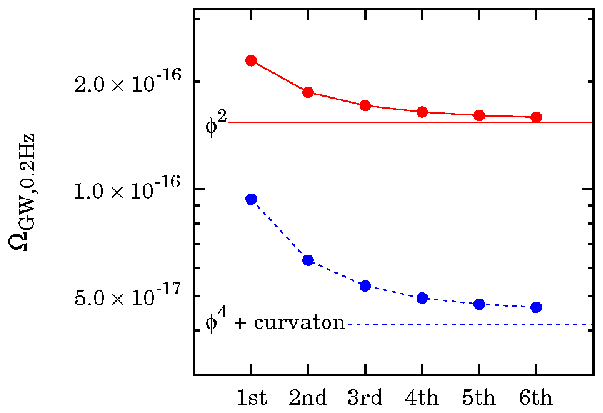}
  \caption{\label{amplitude} Comparison of the overestimation for
    different order truncation.  The vertical axis shows the amplitude
    of the gravitational wave background spectrum $\Omega_{\rm GW}$ at
    the direct detection frequency $f=0.2$Hz.  The points represent
    values calculated by Eq. (\ref{primordial}) truncated at each
    order.  The straight lines correspond to the exact values obtained
    from the numerical calculation.  }

  \includegraphics[width=0.7\textwidth]{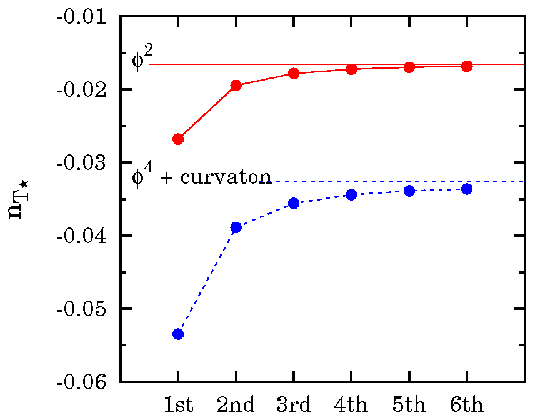}
  \caption{\label{nT} Comparison of the values of $n_T$ estimated with
    Eq. (\ref{primordial}) truncated at different order.  }
 \end{center}
\end{figure}

Table \ref{table1} lists the values of ${\cal P}_T$ at direct
detection frequency ($f=0.2$Hz) for cases of the truncation at each
order in the Taylor expansion.  The degree of overestimation compared
to the numerical result is presented in percentage.  We also list the
values converted to the density parameter of the gravitational wave
background, $\Omega_{\rm GW}\equiv (d\rho_{\rm GW}/d\ln k)/\rho_{c,0}$
\cite{Maggiore:1999vm}, where $\rho_{c,0}\equiv3M_{\rm Pl}^2H_0^2$ is
the critical density of the Universe today and $\rho_{\rm GW}$ is the
energy density of the gravitational waves.  The primordial tensor
spectrum ${\cal P}_T(k)$ can be converted to the present-day density
parameter by using the transfer function as
\begin{equation}
\Omega_{\rm GW}=\frac{1}{12}\left(\frac{k}{H_0}\right)^2 {\cal P}_T(k)T_T^2(k).
\label{OmegaGW1}
\end{equation}
The transfer function is given by
\begin{equation}
T_T^2(k)=(1-\Omega_\Lambda)^2\left(\frac{g_*(T_{\rm hc})}{g_{*0}}\right)\left(\frac{g_{*s0}}{g_{*s}(T_{\rm hc})}\right)^{4/3}
\left(\frac{3}{\sqrt{2}(k\tau_0)^2}\right)^2(1+1.57x_{\rm eq}+3.42x_{\rm eq}^2),
\label{transfer}
\end{equation}
where $\tau_0=2H_0^{-1}$, $x_{\rm eq}=k/k_{\rm eq}$ and $k_{\rm
  eq}\equiv\tau_{\rm eq}^{-1}=7.1\times 10^{-2}\Omega_m h^2 {\rm
  Mpc}^{-1}$ \cite{Kuroyanagi:2009br,Turner:1993vb}.  The effective
number of degrees of freedom is given as $g_*(T_{\rm
  hc})=g_{*s}(T_{\rm hc})=106.75$, when the contribution from the
relativistic standard model particles are taken into account.  The
values at present are $g_{*0}=3.36$ and $g_{*s0}=3.90$.  If we assume
the cosmological parameters to be $\Omega_{\Lambda}=0.734$, $\Omega_m
h^2=0.1334$, $h=0.710$ (taken from the WMAP 7-year mean values
\cite{Komatsu:2010fb}), the amplitude of the primordial spectrum at
the direct detection frequency $f=0.2$~Hz is given by
\begin{equation}
  \Omega_{\rm GW,0.2Hz}=1.36\times 10^{-6}{\cal P}_{T,{\rm 0.2Hz}}.
\label{OGW02Hz}
\end{equation}
Figure \ref{amplitude} plots the amplitude of the present-day tensor
spectrum at direct detection scale ($f=0.2$Hz) in terms of
$\Omega_{\rm GW}$ for different order truncation for the above
mentioned two models.  From the figure, we see that the inclusion of
the higher order terms improves the overestimation significantly.

In figure \ref{nT}, we show how much the overestimation of the
amplitude affects determination of $n_T$.  One may try to determine
the tilt of the spectrum $n_T$ if the amplitude of the gravitational
wave background is determined by both CMB and direct detection
experiments.  However, truncation of the higher order terms would
yield wrong value of $n_T$.  The values listed in table \ref{table1}
and plotted in figure \ref{nT} are estimated with
Eq. (\ref{primordial}) truncating at each order, with the assumption
that ${\cal P}_T$ is determined at both the CMB $k_\star$ and direct
detection scales $k_{\rm 0.2Hz}$.

Below we give some detailed discussion for models considered here: the
quadratic chaotic inflation and a mixed inflaton and curvaton model
with a quartic inflaton potential.

%%%%%%%%%%%%%%%%%%%%
\begin{table}
\begin{center}
\begin{tabular*}{0.7\textwidth}{@{\extracolsep{\fill}}lcccc}
\hline
\hline
 & ${\cal P}_T$ & $\Omega_{\rm GW}$ & overestimation (\%) & $n_{T \star}$\\
\hline
\multicolumn{4}{c}{$\phi^2$model}\\
\hline
numerical & $1.14\times 10^{-10}$ & $1.54\times 10^{-16}$ & 0 & -0.0165\\ 
1st & $1.69\times 10^{-10}$ & $2.30\times 10^{-16}$ & 49 & -0.0268\\ 
2nd & $1.38\times 10^{-10}$ & $1.87\times 10^{-16}$ & 21 & -0.0195\\ 
3rd & $1.27\times 10^{-10}$ & $1.72\times 10^{-16}$ & 11 & -0.0178\\ 
4th & $1.21\times 10^{-10}$ & $1.64\times 10^{-16}$ & 7 & -0.0172\\ 
5th & $1.19\times 10^{-10}$ & $1.61\times 10^{-16}$ & 4 & -0.0170\\ 
6th & $1.17\times 10^{-10}$ & $1.59\times 10^{-16}$ & 3 & -0.0168\\ 

\hline
\multicolumn{4}{c}{$\phi^4$model + curvaton ($r=0.1$)}\\
\hline
numerical & $3.07\times 10^{-11}$ & $4.15\times 10^{-17}$ & 0 & -0.0325\\ 
1st & $6.90\times 10^{-11}$ & $9.35\times 10^{-17}$ & 125 & -0.0535\\ 
2nd & $4.64\times 10^{-11}$ & $6.29\times 10^{-17}$ & 51 & -0.0389\\ 
3rd & $3.93\times 10^{-11}$ & $5.33\times 10^{-17}$ & 28 & -0.0356\\ 
4th & $3.64\times 10^{-11}$ & $4.93\times 10^{-17}$ & 19 & -0.0344\\ 
5th & $3.49\times 10^{-11}$ & $4.74\times 10^{-17}$ & 14 & -0.0339\\ 
6th & $3.42\times 10^{-11}$ & $4.64\times 10^{-17}$ & 12 & -0.0336\\ 
\hline
\hline
\end{tabular*}
\caption{\label{table1} Summary of the amplitude of the primordial
  tensor spectrum ${\cal P}_T$, the present-day density parameter of
  gravitational wave background $\Omega_{\rm GW}$ and percentage of
  overestimation due to the Taylor expansion, which are all evaluated
  at the direct detection frequency $f=0.2$Hz.  The values of $n_{T
    \star}$ evaluated with the truncated expression of the spectrum
  are also listed. }
\end{center}
\end{table}
%%%%%%%%%%%%%%%%%%%%

%%%%%%%%%%%%%%%%%%%%
%%%%%%%%%%%%%%%%%%%%

%%%%%%%%%%%%%%%%%%%%%%%%%%%%%%%%
\subsection{$\phi^2$ model}
%%%%%%%%%%%%%%%%%%%%%%%%%%%%%%%%
In the case of the chaotic inflation with a quadratic potential, 
\begin{equation}
V=\frac{1}{2}m^2\phi^2,
\end{equation}
the slow-roll parameters are given as
\begin{equation}
\begin{array}{rcl}
\epsilon_V&=&\eta_V=2\displaystyle\frac{M_{\rm Pl}^2}{\phi^2},\\
\xi^2_V&=&\sigma^3_V=\tau^4_V=\zeta^5_V=0,
\label{para1_phi2}
\end{array}
\end{equation}
and Eq. (\ref{eqN}) gives
\begin{equation}
N=\frac{\phi^2}{4M_{\rm Pl}^2}-\frac{1}{2}.
\label{N_phi2}
\end{equation}
From Eq. (\ref{eqPS}) we obtain
\begin{equation}
{\cal P}_{S}\simeq \frac{1}{96\pi^2M_{\rm Pl}^6}m^2\phi^4,
\label{eqPSphi2}
\end{equation}
and from Eq. (\ref{eqPT})
\begin{equation}
{\cal P}_{T}\simeq \frac{1}{3\pi^2 M_{\rm Pl}^4}m^2\phi^2.
\label{eqPTphi2}
\end{equation}

The spectrum is calculated assuming that the e-folding number
corresponding to the CMB scale is $N_\star= \ln(a_{\rm
  end}/a_\star)=60$, which gives $\epsilon_{V\star}=8.26\times
10^{-3}$ and $n_{T\star}\simeq -1.65\times 10^{-2}$,
$\alpha_{T\star}\simeq -2.73\times 10^{-4}$, $\beta_{T\star}\simeq
-9.03\times 10^{-6}$, $\gamma_{T\star}\simeq -4.48\times 10^{-7}$,
$\delta_T\simeq -2.96\times 10^{-8}$, $\theta_T\simeq -2.45\times
10^{-9}$.  The mass of the inflaton field $m=1.53\times 10^{13}$GeV is
determined to satisfy the normalization of the scalar perturbations,
${\cal P}_{S\star}=2.43 \times 10^{-9}$, which gives ${\cal
  P}_{T\star}=3.21\times 10^{-10}$.  In this model, the
tensor-to-scalar ratio and the scalar spectral index are $r=0.132$ and
$n_s=0.967$, respectively.

%%%%%%%%%%%%%%%%%%%%%%%%%%%%%%%% 
\subsection{$\phi^4$ model with 
%TT 
the curvaton
}
\label{model_curv}
%%%%%%%%%%%%%%%%%%%%%%%%%%%%%%%%
If we consider the chaotic inflation model with a quartic potential,
\begin{equation}
V=\frac{1}{4}\lambda\phi^4,
\end{equation}
the slow-roll parameters are given as
\begin{equation}
\begin{array}{rcl}
\epsilon_V&=&8\displaystyle\frac{M_{\rm Pl}^2}{\phi^2},\\
\eta_V&=&12\displaystyle\frac{M_{\rm Pl}^2}{\phi^2}=\frac{3}{2}\epsilon_V,\\
\xi^2_V&=&96\displaystyle\frac{M_{\rm Pl}^4}{\phi^4}=\frac{3}{2}\epsilon_V^2,\\
\sigma^3_V&=&384\displaystyle\frac{M_{\rm Pl}^6}{\phi^6}=\frac{3}{4}\epsilon_V^3,\\
\tau^4_V&=&\zeta^5_V=0,
\label{para1_phi4}
\end{array}
\end{equation}
and Eq. (\ref{eqN}) gives 
\begin{equation}
N=\frac{\phi^2}{8M_{\rm Pl}^2}-\frac{3}{2}.
\label{N_phi4}
\end{equation}
We again take the value of the e-folding number as $N_\star=60$, which
gives $\epsilon_{V\star}=1.63\times 10^{-2}$ and $r=0.26$.  This large
tensor-to-scalar ratio is already excluded by current observational
constraints, but it can be avoided by introducing the curvaton
fluctuations.

In the curvaton scenario, the fluctuations in the curvaton field
$\sigma$ produce the scalar perturbations, which results in a
different expression for the tensor-to-scalar ratio
\cite{Langlois:2004nn,Moroi:2005kz,Moroi:2005np,Ichikawa:2008iq}.
Since our interest is in the case where the detection of the
gravitational wave background is possible, we assume here that the
tensor-to-scalar ratio is $r=0.1$.  In this case, fluctuations both
from the inflaton and the curvaton contribute to the primordial
curvature perturbation. In this model, the scalar power spectrum is
given by
\begin{equation}
{\cal P}_S  = {\cal P}_S^{(\phi)} + {\cal P}_S^{(\sigma)} = (1+ \alpha) {\cal P}_S^{(\phi)},
\label{eqPs_curv}
\end{equation}
where ${\cal P}_S^{(\phi)}$ and ${\cal P}_S^{(\sigma)}$ are the
contributions from the inflaton and the curvaton,
respectively. $\alpha$ represents the ratio of the curvaton power
spectrum to the inflaton one at the reference scale, i.e., $\alpha =
{\cal P}_S^{(\sigma)}/{\cal P}_S^{(\phi)}$.  For the $\phi^4$
potential, ${\cal P}_S^{(\phi)}$ is given by
\begin{equation}
{\cal P}_S^{(\phi)} \simeq \frac{1}{768\pi^2M_{\rm Pl}^6}\lambda\phi^6.
\label{eqPs_curv_phi4}
\end{equation}
The scalar spectral index is also modified as
\begin{equation}
n_S=1-2\epsilon_V-\frac{4\epsilon_V-2\eta_V}{1+\alpha}.
\end{equation}

Although the expressions for the scalar perturbation quantities are
modified in this kind of mixed models, the formulae for the tensor
perturbation spectrum ${\cal P}_T$ and the parameters for its scale
dependence, $n_{T}$, $\alpha_{T}$, $\beta_{T}$, $\gamma_{T}$,
$\delta_T$, $\theta_T$, are not modified from the usual inflationary
predictions without the curvaton.  The tensor spectrum in the $\phi^4$
chaotic inflation model is obtained from Eq. (\ref{eqPT}) as
\begin{equation}
{\cal P}_{T}\simeq \frac{1}{6\pi^2 M_{\rm Pl}^4}\lambda\phi^4.
\label{eqPTphi4}
\end{equation}
Since the scalar power spectrum is modified as in Eq.~\eqref{eqPs_curv}, 
the tensor-to-scalar ratio is given by
\begin{equation}
r=\frac{16\epsilon_V}{1+\alpha}.
\end{equation}

Assuming $N_\star=60$, we obtain $n_{T\star}\simeq -3.25\times
10^{-2}$, $\alpha_{T\star}\simeq -5.29\times 10^{-4}$,
$\beta_{T\star}\simeq -1.72\times 10^{-5}$, $\gamma_{T\star}\simeq
-8.39\times 10^{-7}$, $\delta_T\simeq -5.46\times 10^{-8}$ and
$\theta_T\simeq -4.43\times 10^{-9}$.  Given the value
$\epsilon_{V\star}=1.63\times 10^{-2}$, our assumption of $r=0.1$
corresponds to $\alpha_{\star}=1.6$.  The normalization of the scalar
perturbations, ${\cal P}_{S\star}=2.43 \times 10^{-9}$, is used to
determine $\lambda=5.94\times 10^{-14}$, which gives ${\cal
  P}_{T\star}=2.43\times 10^{-10}$.  With this setup, the spectral
index for the scalar perturbation is $n_s=0.961$.

Here we briefly comment on non-Gaussianity in this scenario.  Usually
non-Gaussianity of density fluctuations is represented by so-called
non-linearity parameter $f_{\rm NL}$, which characterizes the size of
3-point function or bispectrum \footnote{
  Current constraints on local- equilateral- and orthogonal-types of
  $f_{\rm NL}$ are (95 \% C.L.) \cite{Komatsu:2010fb}: $ -10 < f_{\rm
    NL}^{\rm local} < 74, -214 < f_{\rm NL}^{\rm equil} < 266$ and $
  -410 < f_{\rm NL}^{\rm equil} < 6 $, respectively.
}.  Since the standard single-field inflation model predicts very
small values of $f_{\rm NL}$ as $f_{\rm NL} \ll \mathcal{O}(1)$, if
the values of $f_{\rm NL}$ is found to be large in the future, it
indicates that we need another source of density fluctuations other
than the inflaton.  As another mechanism of density fluctuations, the
curvaton model \cite{Enqvist:2001zp,Lyth:2001nq,Moroi:2001ct} has been
intensively investigated, and in particular, this model can generate
large non-Gaussianity.  Even if fluctuations from the inflaton also
contribute to the density fluctuations in the Universe, as far as the
curvaton also generates some fraction of the fluctuations, $f_{\rm
  NL}^{\rm local}$ can be large.  Furthermore, large tensor-to-scalar
ratio is also possible in this model, which can be detectable in the
near future.  Note that, when the curvaton is the only source of
density fluctuations, which is usually assumed in many works, the
tensor-to-scalar ratio becomes very small.  However, this kind of
mixed model can give sizable $f_{\rm NL}$ and $r$.

In this mixed scenario where local-type non-Gaussianity is generated,
$f_{\rm NL}$ is given by \cite{Ichikawa:2008iq,Suyama:2010uj}
\begin{equation}
f_{\rm NL}=  \left( \frac{\alpha}{1+\alpha} \right)^2 f_{\rm NL}^{\rm (curvaton)}.
\end{equation}
Here $f_{\rm NL}^{\rm (curvaton)}$ is the one for pure curvaton model
(the curvaton is the only source of density fluctuation).  Depending
on the mass, the decay rate and the initial amplitude of the curvaton
field, $f_{\rm NL}^{\rm (curvaton)}$ can be very large.  Thus, by
tuning these parameters, the case of $\alpha_{\star} = 1.6$ (and
$r=0.1$), which is assumed in this section, can also give large
$f_{\rm NL}$.  Hence, once the gravitational waves and (large)
non-Gaussianity are detected, this kind of scenario would be worth
investigating carefully \cite{Nakayama:2009ce}.

%%%%%%%%%%%%%%%%%%%%%%%%%%%%%%%%
%%%%%%%%%%%%%%%%%%%%%%%%%%%%%%%%
\section{Impact on parameter estimation}
\label{parameterestimate}
%%%%%%%%%%%%%%%%%%%%%%%%%%%%%%%%
%%%%%%%%%%%%%%%%%%%%%%%%%%%%%%%%
Now, in this section, we study the influence of the poor estimation of
the gravitational spectrum amplitude when one adopts the Taylor
approximation truncated at some order.  If direct detection determines
the amplitude of the inflationary gravitational wave background, one
may try to extract information on the inflaton potential and the
e-folding number \cite{Kuroyanagi:2009br} by combining observations of
CMB \cite{Martin:2006rs,Martin:2010kz,Mortonson:2010er} and other
complementary experiments \cite{Barger:2008ii,Adshead:2010mc}.
However, as shown in the previous section, when the power series
expression of the spectrum is adopted, one would overestimate the
amplitude of the gravitational wave spectrum at the direct detection
scale if one truncates the expression at some lower order.  Here we
present how such overestimation of the amplitude affects the
determination of the inflationary parameters by investigating future
constraints.  In this section, we again consider the models discussed
in the previous section.  Note that, in this paper, we do not consider
effect of reheating which may change the shape of the inflationary
gravitational wave background around the direct detection frequency
\cite{Nakayama:2008ip,Nakayama:2008wy,Kuroyanagi:2010mm}.

%%%%%%%%%%%%%%%%%%%%%%%%%%%%%%%%
\subsection{$\phi^2$ model}
%%%%%%%%%%%%%%%%%%%%%%%%%%%%%%%%
If the quadratic chaotic inflation is the model realized in the
nature, the inflationary gravitational wave background could be
directly detected with $\Omega_{\rm GW,0.2Hz}=1.54\times 10^{-16}$
(taken from the numerical result, given in Table \ref{table1}), which
is obtained assuming $N_\star =60$ and the scalar perturbation
being normalized as ${\cal P}_{S\star}=2.43 \times 10^{-9}$.  With the
power-law approximation, one can describe the amplitude of the
gravitational wave background at direct detection scale as
\begin{eqnarray}
\Omega_{{\rm GW,0.2Hz}}
=1.36\times 10^{-6}{\cal P}_{T\star}
\exp[-2(38.7\epsilon_{V\star})
-\frac{4}{2!}(38.7\epsilon_{V\star})^2-\frac{16}{3!}(38.7\epsilon_{V\star})^3\nonumber\\
-\frac{96}{4!}(38.7\epsilon_{V\star})^4-\frac{768}{5!}(38.7\epsilon_{V\star})^5-\frac{7680}{6!}(38.7\epsilon_{V\star})^6+\cdots],
\label{omegaGW_phi2}
\end{eqnarray}
where we have used Eqs. (\ref{primordial}), (\ref{coeff}),
(\ref{OGW02Hz}), (\ref{para1_phi2}) and $\ln(k_{0.2{\rm
    Hz}}/k_\star)=38.7$.  Notice that, from the above expression, the
relation between ${\cal P}_{T\star}$ and $\epsilon_{V\star}$ can be
provided once the value of $\Omega_{\rm GW,0.2Hz}$ is determined.  The
values of ${\cal P}_{T\star}$ and $\epsilon_{V\star}$ directly give
information on the e-folding number $N_\star$ and the mass of the
inflaton $m$ via the following relations,
\begin{equation}
N_\star=\frac{1}{2\epsilon_{V\star}}-\frac{1}{2},
\label{Nstar_phi2}
\end{equation}
\begin{equation}
m^2=\frac{3\pi^2M_{\rm Pl}^2\epsilon_{V\star}}{2}{\cal P}_{T\star},
\label{m_phi2}
\end{equation}
which follow from Eqs. (\ref{N_phi2}) and (\ref{eqPTphi2}).

\begin{figure}
 \begin{center}
  \includegraphics[width=0.4\textwidth]{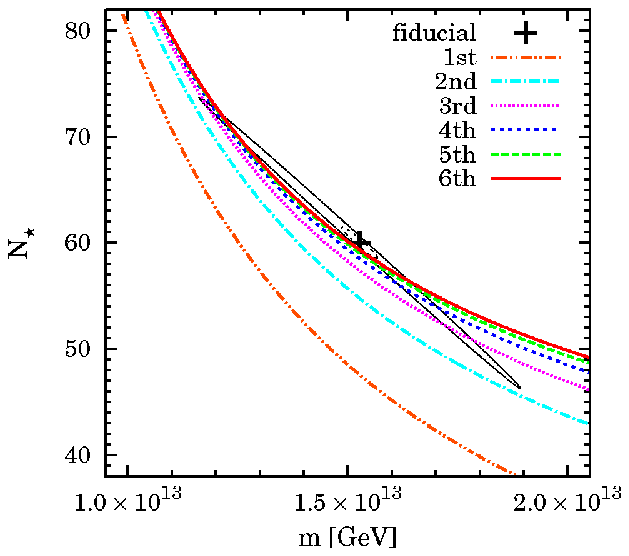}
  \caption{\label{m_N1} Parameter estimation for the $\phi^2$ model.
    The values of $m$ and $N_\star$ are inferred from direct detection
    of the inflationary gravitational wave background with
    $\Omega_{\rm GW,0.2Hz}=1.54\times 10^{-16}$.  Each line represents
    the values derived assuming the gravitational wave spectrum is
    described by Eq. (\ref{omegaGW_phi2}), truncated at first, second,
    third, fourth, fifth and sixth order, respectively.  The fiducial
    point is shown as a cross mark.  The ellipses are the marginalized
    $2\sigma$ constraints expected from Planck (solid) and CMBpol
    (dashed).}
 \end{center}

 \begin{center}
  \includegraphics[width=0.98\textwidth]{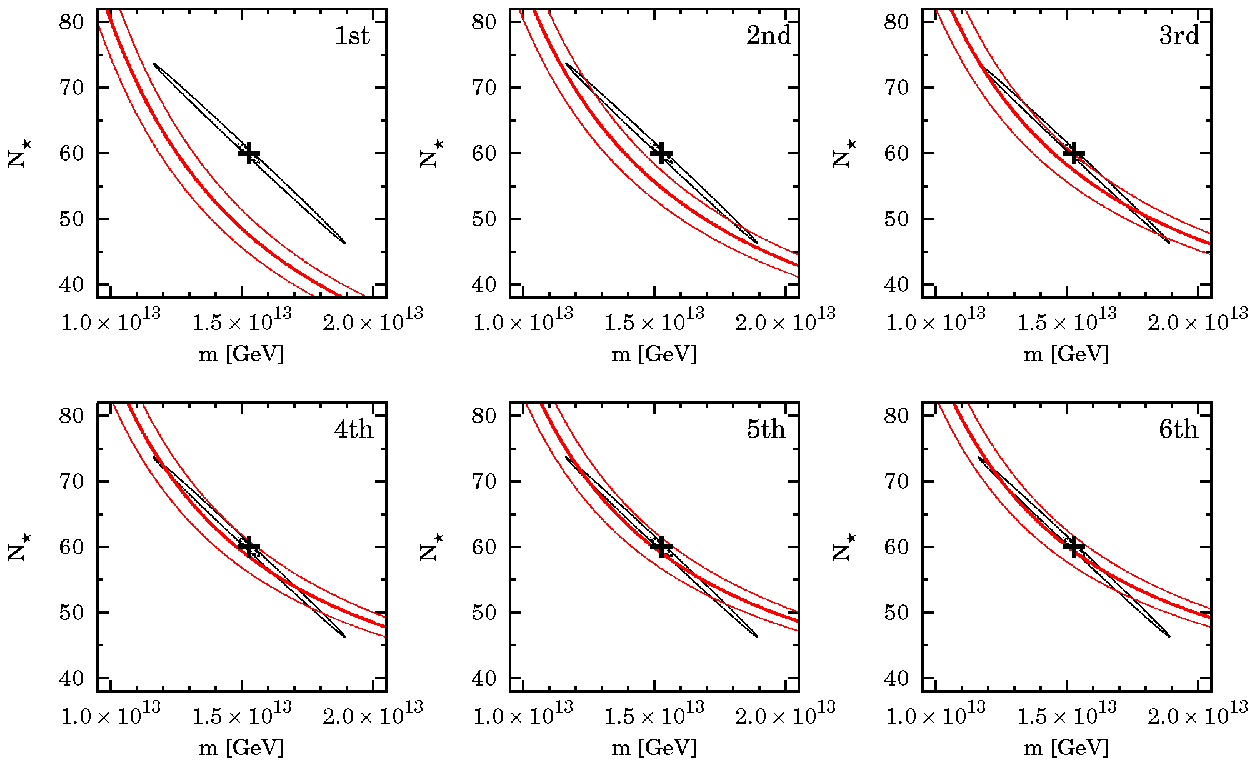}
  \caption{\label{m_N2} The values of $m$ and $N_\star$ inferred from
    the determination of $\Omega_{\rm GW,0.2Hz}$ with the $2\sigma$
    experimental error of DECIGO/BBO.  Each panel is for a different
    order truncation.}
 \end{center}
\end{figure}

In Figure \ref{m_N1}, we show parameter constraints expected from
future CMB observations in the $m-N_\star$ plane as well as the the
values of $m$ and $N_\star$ which give $\Omega_{\rm
  GW,0.2Hz}=1.54\times 10^{-16}$ at the direct detection scale for
several cases of the truncation in Taylor expansion at some order.
Expected CMB constraints are derived from the Fisher matrix analysis
\cite{CMBp1,CMBp2} with the instrumental sensitivity of Planck
\cite{:2006uk} and CMBpol \cite{Baumann:2008aq}, taking into account
the analysis of both temperature and polarization data up to the
multipole $l=2000$.  The uncertainties on $m$ and $N_\star$ are
obtained by transforming parameters from $(n_S,r,{\cal P}_{S\star})$
into $(m, N_\star)$ \cite{Kuroyanagi:2009br}, with other cosmological
parameters $(h,\Omega_bh^2,\Omega_ch^2,\tau)=(0.710,
0.1109,0.02258,0.088)$ \cite{Komatsu:2010fb} marginalized over.

Figure \ref{m_N1} illustrates an important fact that the values of $m$
and $N_\star$ are estimated incorrectly when one determines the
parameters from direct detection experiments using the power series
expression with higher order terms being neglected.  The lines in the
$m-N_\star$ plane are plotted by Eqs. (\ref{Nstar_phi2}) and
(\ref{m_phi2}) with parameters ${\cal P}_{T\star}$ and
$\epsilon_{V\star}$ satisfying Eq. (\ref{omegaGW_phi2}), truncated at
each order.  The fiducial values of $m$ and $N_\star$ are taken to be
the same as in Sec. \ref{overestimation}.  Neglect of the higher order
terms leads to an underestimation of ${\cal P}_{T\star}$ or an
overestimation of $\epsilon_{V\star}$, which results in an incorrect
estimation of the values of $m$ and $N_\star$.  As seen from the
figure, the deviation of the line from the true (fiducial) value
becomes larger as the power series expansion is truncated at lower
order.

In particular, for the case of truncation at first or second order,
the deviation is not negligible even if the error in measuring
$\Omega_{\rm GW}$ is taken into account.  To present this clearly, we
also plot the expected error in future direct detection experiments in
Fig.~\ref{m_N2}.  We assume that future experiments determine the
value of $\Omega_{\rm GW}$ with an accuracy of
\cite{Seto:2005qy,Kudoh:2005as,Corbin:2005ny}
\begin{equation}
\sigma_{\Omega_{\rm GW}}=8.0\times 10^{-18}\left(\frac{10^{-16}}{\Omega_{\rm GW}}\right),
\label{sigmaOGW}
\end{equation}
which is derived from the sensitivity of the BBO experiment (Detailed
values for computing the noise spectrum are given in
Ref. \cite{Kuroyanagi:2010mm}).  The Fabry-Perot type DECIGO has a
similar sensitivity.  Thus, the region within the error band would be
similar to parameter space allowed by constraints from direct
detection by DECIGO or BBO.  Therefore, the use of power series
expression of the spectrum may lead to incorrect parameter constraints
from direct detection experiments, when one truncates it at lower
order.  However, if we includes up to the sixth order term, the
estimate almost coincides with the true value.

%%%%%%%%%%%%%%%%%%%%
%%%%%%%%%%%%%%%%%%%%

%%%%%%%%%%%%%%%%%%%%%%%%%%%%%%%%
\subsection{$\phi^4$ model with the curvaton}
%%%%%%%%%%%%%%%%%%%%%%%%%%%%%%%%
Next, we show an example of parameter estimation for the quartic
potential in the presence of the contribution from the curvaton
fluctuations to the primordial scalar perturbations.  In the same way
as in Sec. \ref{model_curv}, we assume the tensor-to-scalar ratio to
be $r=0.1$.  In this case, the amplitude of the gravitational wave
background would be determined to be $\Omega_{\rm GW,0.2Hz}=4.15\times
10^{-17}$ by direct detection experiments.  The determination of
$\Omega_{\rm GW,0.2{\rm Hz}}$ provides a relation between ${\cal
  P}_{T\star}$ and $\epsilon_{V\star}$ via
\begin{eqnarray}
\Omega_{{\rm GW},0.2{\rm Hz}}
=1.36\times 10^{-6}{\cal P}_{T\star}
\exp[-2(38.7\epsilon_{V\star})
-\frac{2}{2!}(38.7\epsilon_{V\star})^2
-\frac{6}{3!}(38.7\epsilon_{V\star})^3\nonumber\\
-\frac{12}{4!}(38.7\epsilon_{V\star})^4
-\frac{48}{5!}(38.7\epsilon_{V\star})^5
-\frac{240}{6!}(38.7\epsilon_{V\star})^6+\cdots],
\label{omegaGW_phi4}
\end{eqnarray}
where we have used Eqs. (\ref{primordial}), (\ref{coeff}),
(\ref{OGW02Hz}), (\ref{para1_phi4}) and $\ln(k_{0.2{\rm
    Hz}}/k_\star)=38.7$.  It can be converted to the information on
$N_\star$ and $\lambda$ by
\begin{equation}
N_\star=\frac{1}{\epsilon_{V\star}}-1,
\label{Nstar_phi4}
\end{equation}
\begin{equation}
\lambda=\frac{3\pi^2\epsilon_{V\star}^2}{32}{\cal P}_{T\star},
\label{lambda_phi4}
\end{equation}
which follows Eqs. (\ref{N_phi4}) and (\ref{eqPTphi4}).

\begin{figure}
 \begin{center}
  \includegraphics[width=0.4\textwidth]{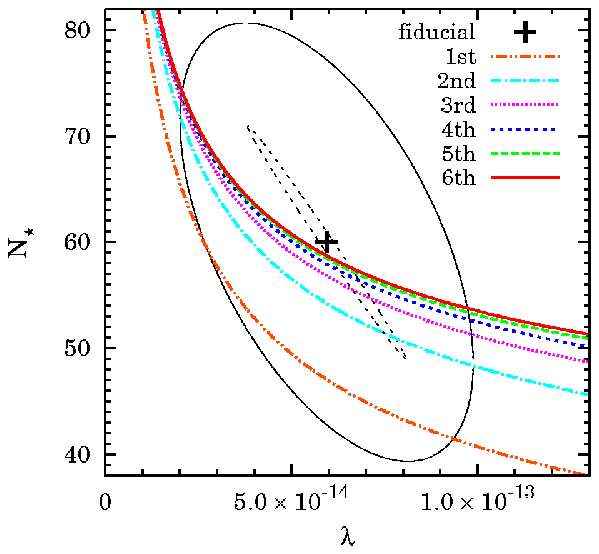}
  \caption{\label{l_Ncurv1} Parameter estimation for the $\phi^4$
    model with the curvaton.  The tensor-to-scalar ratio is assumed to
    be $r=0.1$, which corresponds to direct detection with
    $\Omega_{\rm GW,0.2Hz}=4.15\times 10^{-17}$. Each line represents
    the values derived assuming the gravitational wave spectrum is
    described by Eq. (\ref{omegaGW_phi4}), truncated at first, second,
    third, fourth and sixth order, respectively.  The ellipses are the
    marginalized $2\sigma$ constraints expected from Planck (solid)
    and CMBpol (dashed).}
 \end{center}

 \begin{center}
  \includegraphics[width=0.98\textwidth]{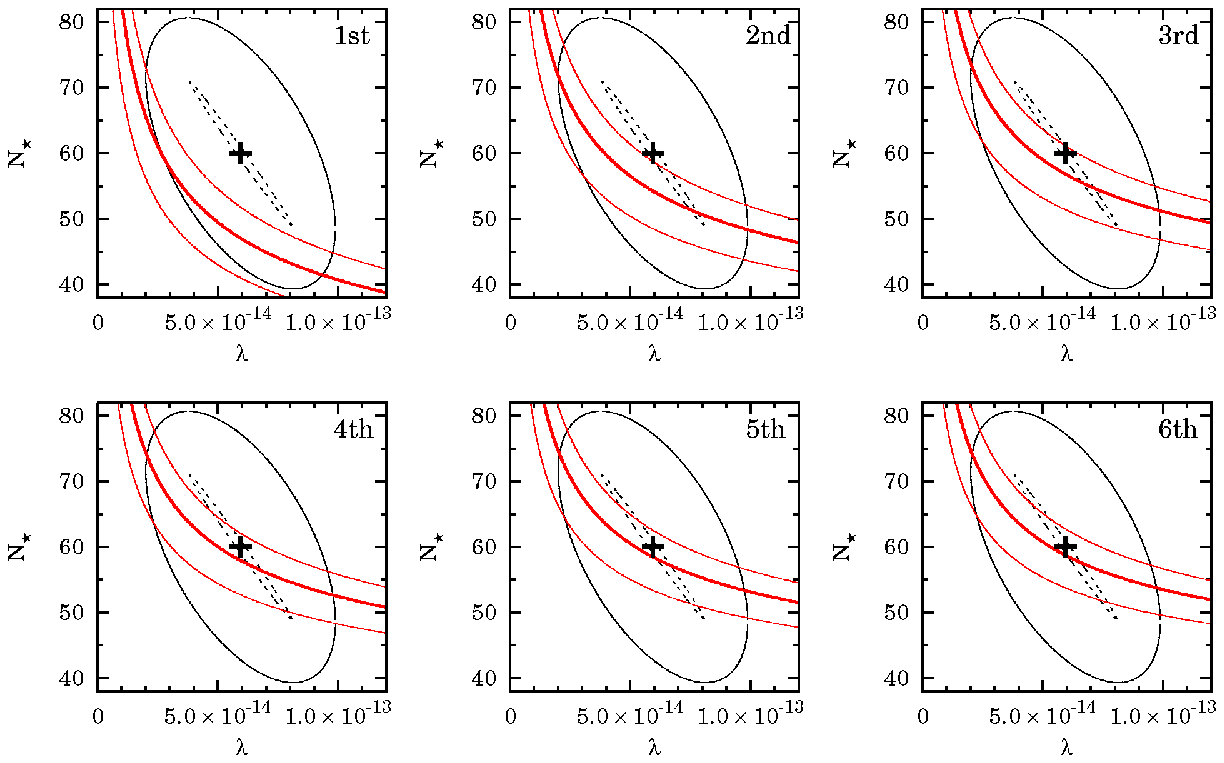}
  \caption{\label{l_Ncurv2} The values of $\lambda$ and $N_\star$
    inferred from the determination of $\Omega_{\rm GW,0.2Hz}$ with
    the $2\sigma$ experimental error of DECIGO/BBO.  Each panel is for
    a different order truncation.}
 \end{center}
\end{figure}

In Fig. \ref{l_Ncurv1}, the values of $\lambda$ and $N_\star$ obtained
from the determination of $\Omega_{{\rm GW},0.2{\rm Hz}}$ are shown
for different order truncation of Eq. (\ref{omegaGW_phi4}).  The lines
in the $\lambda-N_\star$ plane are plotted by Eqs. (\ref{Nstar_phi4})
and (\ref{lambda_phi4}) with parameters ${\cal P}_{T\star}$ and
$\epsilon_{V\star}$ satisfying Eq. (\ref{omegaGW_phi4}).  For the same
reason as described in the previous subsection, the truncation of the
higher order terms gives an incorrect estimation of the values of
$\lambda$ and $N_\star$.  The deviation from the true value is larger
than the $\phi^2$ case, because of the larger overestimation of the
spectrum as presented in Sec. \ref{overestimation}.  In this case, the
overestimation may come not only from the truncation of the higher
order terms of the power series expansion in terms of $\ln k$, but
also those of the slow-roll approximation.  In our numerical
calculation, the slow-roll parameter is $\epsilon_V\simeq 4.56\times
10^{-2}$ when the mode corresponding to $0.2$~Hz exits the horizon
during inflation.  This means the second order slow-roll correction in
${\cal P}_{T, 0.2{\rm Hz}}$ (see Eq. (\ref{eqPT2nd})) can be a few
percent around direct detection scales.  Note that this cannot be
improved even if we take into account the second order slow-roll
correction as long as the spectrum is extrapolated from CMB scales,
since the second order slow-roll correction is still small
($\epsilon_{V\star}\simeq 1.63\times 10^{-2}$) when the modes
corresponding to CMB scales exit the horizon.

In Fig.  \ref{l_Ncurv2}, the lines are shown with the experimental
error of direct detection experiments, estimated by
Eq. (\ref{sigmaOGW}).  The larger error than the $\phi^2$ case is
because of the smaller amplitude of the tensor spectrum due to the
reduced normalization.  Furthermore, since the CMB constraints are
obtained marginalizing over not only the cosmological parameters but
also $\alpha_\star$ characterizing the contribution of the curvaton
fluctuations, the uncertainty becomes larger compared to the case for
the $\phi^2$ chaotic inflation model without the curvaton.

%%%%%%%%%%%%%%%%%%%%
%%%%%%%%%%%%%%%%%%%%

%%%%%%%%%%%%%%%%%%%%%%%%%%%%%%%%
%%%%%%%%%%%%%%%%%%%%%%%%%%%%%%%%
\section{Conclusion}
\label{conclusion}
%%%%%%%%%%%%%%%%%%%%%%%%%%%%%%%%
%%%%%%%%%%%%%%%%%%%%%%%%%%%%%%%%
Inflation robustly predicts a stochastic gravitational wave background
with a nearly scale-invariant spectrum.  The detection of such
gravitational waves is one of the next challenges in observational
cosmology.  If both CMB polarization and direct detection experiments
achieve the detection, the independent information from the two
different scales would provide a breakthrough in understanding the
underlying physics of inflation.

Since the two different experiments measure gravitational waves at
wavelengths separated by 16 orders of magnitude, the deviation from
the scale-invariant spectrum, which is traditionally expressed by the
power series expansion of $\ln k$, causes a large difference in
amplitude of the primordial spectrum between two scales.  The
difference comes not only from the first order term of the power-law
expansion, so-called the spectral index, but also from higher order
terms, so-called runnings.  We have presented that, in the case of
chaotic inflation, the truncation of the running terms leads to the
overestimation of the spectrum amplitude at the direct detection
frequency.  The overestimation is more prominent in the case where
inflation predicts large slow-roll parameters.  If we consider a
single-field inflation model, large slow-roll parameters correspond to
a large tensor-to-scalar ratio, in case of which we expect to detect
the inflationary gravitational waves.  Therefore, the overestimation
of the tensor power spectrum should be carefully taken into
consideration in case we achieve detection of the inflationary
gravitational wave background.

Furthermore, we have investigated how the overestimation affects the
determination of inflationary parameters.  We have considered
parameter constraints obtainable from future direct detection
experiments, assuming a specific form of the inflation potential.  We
have presented two examples: a quadratic chaotic inflation model and
mixed inflation and curvaton model with a quartic inflaton potential.
In both cases, the use of truncated power-law spectrum causes an
incorrect estimation of the inflationary parameters and it can be
improved by adding higher order terms.  For correct estimation of
inflationary parameters, we need to take into account higher order
terms, perform a numerical calculation, or develop a new
parametrization of the spectrum to connect the two separate scales.

\acknowledgments{ The authors are grateful to Takeshi Chiba for
  helpful comments.  SK would like to thank Toyokazu Sekiguchi,
  Takeshi Kobayashi and Shuichiro Yokoyama for discussions.  TT would
  like to thank Kari Enqvist for discussions.  The work of TT is
  partially supported by the Grant-in-Aid for Scientific research from
  the Ministry of Education, Science, Sports, and Culture, Japan,
  No.\,23740195 and Saga University Dean's Grant 2011 For Promising
  Young Researchers.  }

%%%%%%%%%%%%%%%%%%%%%%%%%%%
%%%%%%%%%%%%%%%%%%%%%%%%%%%

\end{document}